\documentclass[debug]{rmaa}

\title{Radial abundance gradients from planetary nebulae at different
distances from the galactic plane} 

\author{
  W. J. Maciel,\altaffilmark{1} 
  R. D. D. Costa,\altaffilmark{1}
  and O. Cavichia\altaffilmark{2}}

\altaffiltext{1}{Instituto de Astronomia, Geof\'\i sica e Ci\^encias
                 Atmosf\'ericas, Universidade de S\~ao Paulo, Brazil.}
\altaffiltext{2}{Instituto de F\'\i sica e Qu\'\i mica, Universidade 
                 Federal de Itajub\'a, Brazil.}

\shortauthor{Maciel, Costa \& Cavichia}
\shorttitle{Radial abundance gradient variations}

\fulladdresses{
\item W. J. Maciel, and R. D. D. Costa: 
    Instituto de Astronomia, Geof\'\i sica e Ci\^encias
    Atmosf\'ericas, Universidade de S\~ao Paulo - 
    Rua do Mat\~ao 1226, CEP 05508-090, S\~ao Paulo SP, Brazil
    (wjmaciel@iag.usp.br, roberto.costa@iag.usp.br.)

\item O. Cavichia: 
     Instituto de F\'\i sica e Qu\'\i mica, Universidade Federal 
     de Itajub\'a, Av. BPS 1303, Pinheirinho, CEP 37500-903, 
     Itajub\'a, MG, Brazil 
    (cavichia@unifei.edu.br.)}

\listofauthors{W. J. Maciel, R. D. D. Costa & O. Cavichia}
\indexauthor{Maciel, W. J.}
\indexauthor{Costa, R. D. D.}
\indexauthor{Cavichia, O.}

\abstract{
We investigate the variations of the radial O/H abundance gradients 
from planetary nebulae (PN) located at different distances from the 
galactic plane. In particular, we determine the abundance 
gradients at different heights from the plane in order to
investigate a possible gradient inversion for the objects at larger 
distances from the plane. We consider a large sample of PN with known 
distances, so that the height relative to the galactic plane can be 
derived, and accurate abundances, so that the gradients can be determined. 
}

\resumen{ }

\addkeyword{Galaxies: Abundances}
\addkeyword{Galaxies: Disk}
\addkeyword{ISM: abundances}
\addkeyword{ISM: planetary nebulae: general}

\begin{document}
\maketitle

\section{Introduction}
\label{section1}

There are some evidences that the galactic radial abundance gradient from elements such as Fe, O, 
etc. changes according to the ages of the objects considered. Two main modifications have 
been suggested in the literature: first, the average abundances decrease as older objects 
are considered, and the gradient itself may or may not be affected. Second, there are some 
suggestions that in older objects there is an inversion of the gradients at low galactocentric 
distances, $R < 8\,$kpc, approximately, in the sense that the decreasing outwards abundances 
also decrease inwards, as older objects are considered (see for example Sancho-Miranda et al. 
\citeyear{sancho}, Cheng et al. \citeyear{cheng1}, \citeyear{cheng2}, Carrell et al. 
\citeyear{carrell}). 

The results by Cheng et al. (\citeyear{cheng1}, \citeyear{cheng2}) suggest that the radial 
gradients flatten out at large distances from the plane, which is confirmed by a sample of 
dwarf stars from the SEGUE survey at $7 < R {\rm (kpc)} < 10.5$ with proper motions and 
spectroscopic abundances by Carrell et al. (\citeyear{carrell}).  The latter includes objects 
up to $\vert z \vert \simeq 3\,$kpc, for which slightly positive gradient are obtained for the 
[Fe/H] ratio. Other previous investigations for thick disk stars also led to flat or slightly 
positive gradients (Allende-Prieto et al. \citeyear{allende}, Nordstr\"om et al. 
\citeyear{nordstrom}).

Some recent chemical evolution models predict a strong flattening of the [Fe/H] radial 
gradient for higher distances from the galactic plane, $\vert z \vert  > 0.5\,$kpc, as 
shown for example by Minchev et al. (\citeyear{minchev}, see also Chiappini et al. 
\citeyear{chiappini} and references therein). These models take into account radial 
migration, and suggest an inversion of the gradient, from negative to weakly positive, 
for $R < 10\,$kpc  and $0.5 < \vert z \vert < 1.0\,$kpc. Also, theoretical models  by 
Curir et al. (\citeyear{curir}) and Spitoni \& Matteucci (\citeyear{spitoni}) predict a 
gradient inversion at galactocentric distances of about 10 kpc for the early Galaxy 
(about 2 Gyr), which is a consequence of the inside-out formation of the thick disk.

Differences in the radial abundance gradients are also supported by Fe/H and Si/H 
gradients as derived for red clump stars from the RAVE survey, in the sense that the 
steeper gradients occur for low distances to the plane, becoming flatter for higher  
$z$ values.  A slightly positive gradient has been found from APOGEE data for stars at 
$1.5 < z {\rm (kpc)} < 3.0$ (Chiappini et al. \citeyear{chiappini}, Boeche et al. 
\citeyear{boeche}).

Planetary nebulae may give some contribution to this problem. The nebulae in the disk of
the Galaxy are spread along at least 1 kpc above (or below) the galactic plane, that is, these 
objects have typically $0 < \vert z \vert  {\rm (pc)} < 1,000$. In view of the number of 
PN with known abundances, which is of the order of about 300 objects, it is possible that 
considering objects at different heights from the plane, some differences between the 
derived gradients may be determined. In principle, PN located higher from the galactic 
plane are produced by older stars, as in the case of  the halo PN, which are very probably 
older than the disk nebulae.

In this work, we investigate the possible variations in the radial abundance gradients
when objects at different distances from the galactic plane are taken into account. We 
examine a large sample of galactic PN with known distances and accurate abundances, so that 
both the height $z$ relative to the galactic plane and the galactocentric distance $R$ can 
be determined, apart from the radial abundance gradients. We consider the oxygen abundance 
ratio O/H, which has been derived for a large number of objects and present the smallest 
uncertainties in the abundances.

\section{The Data}
\label{section2}

The main problem in studying gradients from planetary nebulae is to obtain their distances. 
In this paper, we will consider the objects from the Magellanic Cloud calibration by 
Stanghellini et al. (\citeyear{stanghellini}) and, in particular from Stanghellini and Haywood 
(\citeyear{SH}, hereafter SH), who also give chemical abundances. This is the most recent 
sizable distance scale in the literature, so that the probability of obtaining a reasonable 
estimate of the gradients is higher. Table~1 of Stanghellini and Haywood (\citeyear{SH}) 
includes 728 objects, which will be our starting sample. The objects are referenced by their 
PNG number, to which we added the common names from the catalogue by Acker et al. 
(\citeyear{acker}). Table~2 of Stanghellini and Haywood (\citeyear{SH}) gives nebular abundances 
of He, N, O, and Ne for a smaller sample, containing 224 objects. Since the galactic coordinates 
$(l,b)$ are known, by taking the distances into account, the $z$ coordinate can be obtained, as 
well as the galactocentric distance $R$, adopting $R_0 = 8\,$kpc for the distance of the Sun to  
the galactic centre. As an alternative to the abundances given by SH, we will also consider 
the abundances by the IAG group as discussed for example by Maciel and Costa  (\citeyear{mc2013}).
This includes 234 disk nebulae. Considering the objects of both samples as part of 
the 728 PN sample given by SH, oxygen abundances are known for 201 objects from the SH sample; 
222 objects from the IAG sample; 160 objects in common for the SH and IAG samples, and 263 objects 
including all objects with abundances in at least one of the two samples. This is the initial 
sample of galactic disk planetary nebulae that we will consider in this paper. 

The reason why we can safely consider both samples (SH and IAG) concerning their oxygen chemical 
abundances can be seen in Figure~1, where we plot the SH oxygen abundances ($y$ axis) as a 
function of the IAG abundances ($x$ axis). A linear fit of the form $y = a\, x$ gives the 
following results: $a = 0.9898$, with a standard deviation $\sigma = 0.0019$, and correlation 
coefficient $r = 0.9997$, with an uncertainty $\sigma_r = 0.2096$. The least squares fit line 
(dashed line) is very close to the one-to-one line (full line), only slightly displaced 
downwards by about 0.05 dex. It can be seen that only a few nebulae have higher deviations 
than the average.

   \begin{figure}
   \centering
   \includegraphics[angle=0, width=11.0cm]{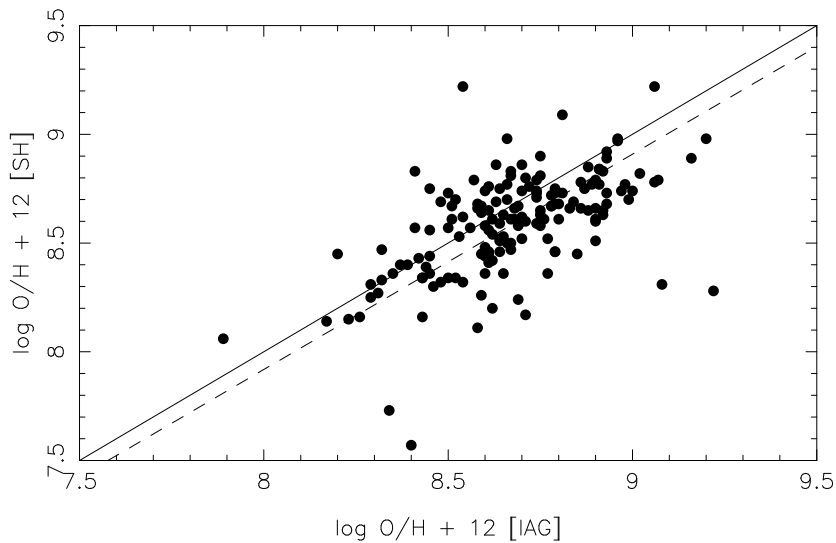}
   \caption{Comparison of the oxygen abundances by Stanghellini \& Haywood (2010) (y axis) 
   and the abundances by the IAG sample (x axis).}
   \label{fig1}
   \end{figure}

\section{The Method}
\label{section3}

In order to investigate the possible variations of the abundance gradient for older objects, 
which are in principle located higher above the galactic plane, we have divided the 263 object 
sample into several groups, according to the $z$ values, which were then compared with the 
results for the whole sample. Once the groups have been defined, the procedure consists in 
plotting $\epsilon{\rm(O)} = \log {\rm O/H} + 12$ as a function of the galactocentric distance 
$R$, and estimating the average linear gradient, as defined by the equation

$$\epsilon{\rm(O)} = \log {\rm O/H} + 12 = A + B\ R \eqno(1)$$

\noindent
and then estimating the coefficients $A$ and $B$, their uncertainties and the correlation 
coefficient $r$. Alternatively, we are also interested in investigating any possible variation 
of the slope $B$ along the galactocentric radius, so that we will also obtain polynomial fits 
using a second degree function defined as

$$\epsilon{\rm(O)} = C + D\ R + E\ R^2 \eqno(2)$$

\noindent
and then estimating the coefficients $C$, $D$, $E$, and the $\chi$ square value. We have  
considered two cases: Case A, in which we have adopted the abundances by SH if both samples 
present abundances for the same object, and Case B, in which abundances from the IAG sample 
were selected in this case. As it turned out, both Case A and B produce essentially the same 
results, so that most figures presented here are for Case A.

\section{Results and Discussion}
\label{section4}

\subsection{The Total Disk Sample}

The first results are obtained for the whole sample of 263 objects, both for Case A and B.
The derived linear fits defined as in equation (1) are shown in the first two lines
of Table~1. The corresponding figures are shown in Figure~2, which extends up to 15 kpc 
from the centre, so that the few objects at higher galactocentric distances are not included. 
It can be seen that in both cases an average gradient  $d{\rm (O/H)}/dR \simeq -0.02$ to 
$-0.03$ dex/kpc is obtained. The average uncertainty in the abundances are about $-0.2$ dex. 
The dispersion is larger at larger galactocentric distances, as expected, and the correlation 
coefficient is relatively low, which is a consequence of the relatively flat gradient and the
uncertainties in the abundances.

   \begin{figure}
   \centering
   \includegraphics[angle=0, width=11.0cm]{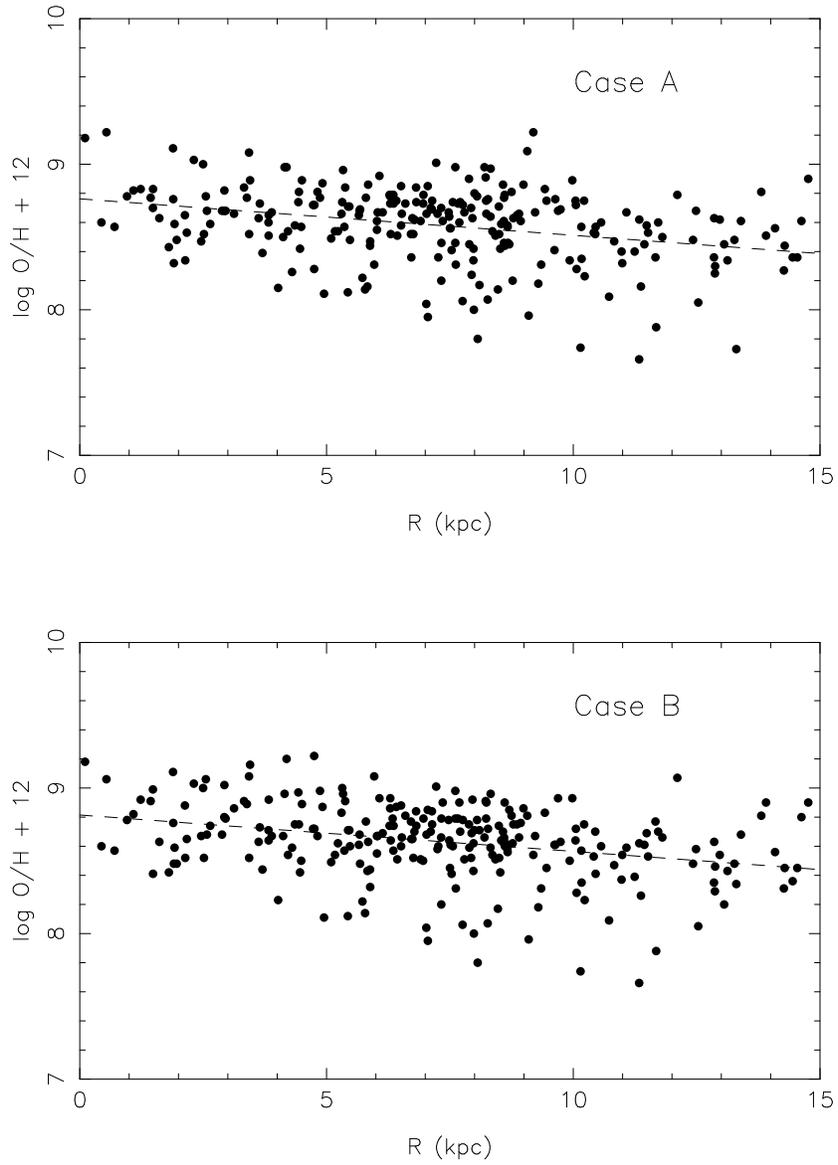}
   \caption{ Variation of the oxygen abundance with the galactocentric distance for the 
    whole sample, Cases A and B.}
   \label{fig2}
   \end{figure}

\begin{table*}
\small
\caption[]{Coefficients of the linear fits given by equation (1).}
\label{table1}
\begin{flushleft}
\begin{tabular}{ccccccc}
\noalign{\smallskip}
\hline\noalign{\smallskip}
Case & $A$  & $\sigma_A$ & $B$ & $\sigma_B$  & $r$ & $\sigma_r$ \\
\noalign{\smallskip}
\hline\noalign{\smallskip}
all data & & & & & & \\
A  &  8.7625  &  0.0349  &  $-$0.0251  &  0.0041  &  $-$0.3565  &  0.2529 \\
B  &  8.8146  &  0.0342  &  $-$0.0250  &  0.0040  &  $-$0.3626  &  0.2474 \\
& & & & & & \\
binned data & & & & & & \\
A  &  8.7390  &  0.0316  &  $-$0.0223  & 0.0029  &  $-$0.9443  &  0.0487  \\	
B  &  8.7885  &  0.0344  &  $-$0.0213 &  0.0032  &  $-$0.9289  &  0.0532  \\	
& & & & & & \\
disk + bulge nebulae & & & & & & \\
A  &  8.6596  &  0.0265  &  $-$0.0156  &  0.0035  &  $-$0.2292  &  0.2731 \\
B  &  8.6780  &  0.0266  &  $-$0.0124  &  0.0035  &  $-$0.1828  &  0.2743 \\
\noalign{\smallskip}
\hline
\end{tabular}
\end{flushleft}
\end{table*}

We can probably get a more meaningful result by dividing the sample according to their 
galactocentric distances, for example using 2 kpc bins. We have considered 9 bins, 
and the results are shown in Figure~3,  where the data points are shown as empty circles 
and the average abundances in each bin as filled circles. The vertical error bars show the mean 
deviation and the horizontal bars show the adopted galactocentric limits. The dashed line 
shows the linear fits as given by lines 3 and 4 of Table~1. 

   \begin{figure}
   \centering
   \includegraphics[angle=0, width=11.0cm]{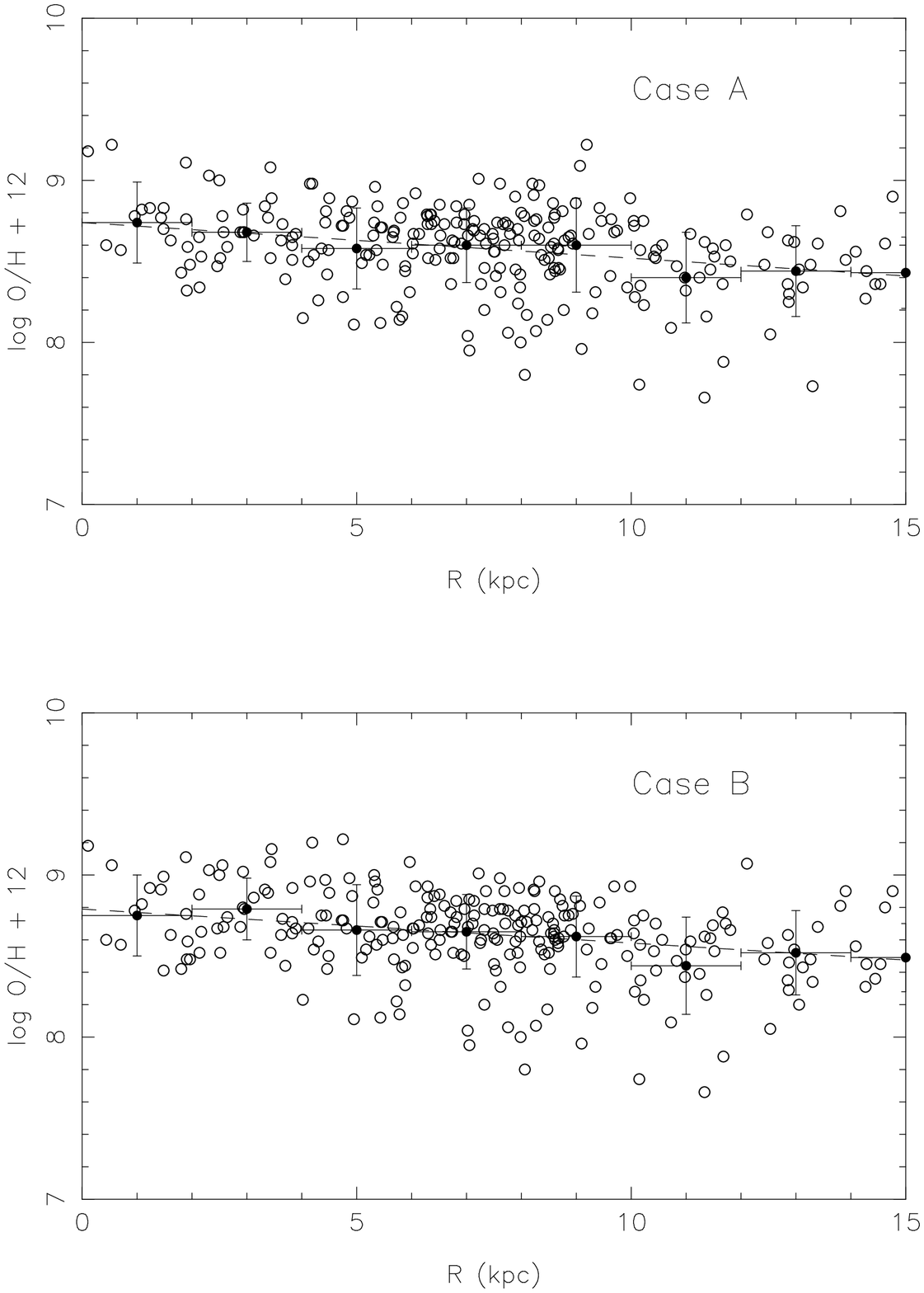}
   \caption{The same as Figure~2, adopting average abundances in 2 kpc bins.}
   \label{fig3}
   \end{figure}

A comparison of the dashed lines in figures~2 and 3 show that they are very similar, that is, 
the linear fits of the whole sample and of the binned data are essentially the same. This can 
also be seen from Table~1, which shows that the intercept and slope of both lines are very 
similar. However, the correlation coefficients of the binned data are much higher than for 
the whole sample.

A more detailed analysis can be made obtaining a polynomial fit to the data, so that any 
slope variation with position can be estimated. We have considered a second order polynomial 
fit to the data, as defined in equation (2).  The obtained coefficients are
$C = 8.7581$, $D = -0.0238$, and $E = -0.0001$ for Case A, with $\chi =	0.0643$,
and $C = 8.8364$, $D = -0.0312$, and $E = 0.0004$ with $\chi = 0.0614$ for Case B.
In Case A the linear and quadratic fits are almost the same, and for Case B there is a 
slight difference between them for $R > 10\,$kpc. Also, for Case A the gradients are essentially 
constant for the whole range of galactocentric distances, while for Case B some flattening 
is observed at large R. The gradient varies from  $d{\rm (O/H)}/dR \simeq -0.024$ to 
$-0.027\,$dex/kpc for Case A, with similar results for Case B.

Considering the total sample of disk objects, it can be seen that the average 
gradients found here are similar or somewhat flatter than previous determinations 
(see for example Maciel \& Costa \citeyear{mc2013}). Since PN are relatively old objects, 
compared for example with HII regions, they may have been displaced from their original 
birthplaces. As a consequence, the measured gradients are probably flatter than before 
the displacement occurred. In other words, gradients derived from planetary nebulae are 
probably  a lower limit to the original gradient. In fact, simulations with a large number 
of objects show that there is no way the gradients can be increased by the displacement of 
the progenitor stars. Additionally, most PN samples  considered so far in the literature 
include objects of very different ages, as we have shown  in our previous work on the age 
determination of the PN progenitor stars (Maciel et al. \citeyear{mci2010}, 
\citeyear{mrc2011}, Maciel \& Costa \citeyear{mc2013}), which also contributes to the
flattening of the gradients.

Some recent chemodynamical models (see for example Chiappini et al. \citeyear{chiappini}, 
and references  therein) suggest that the solar neighbourhood has been contaminated with 
stars born at lower galactocentric distances, which are more metal-rich, since they come 
from a more metal-rich environment. As a result, a determination of the gradient on the 
basis of relatively old stars, such as the PN progenitor stars, lead to a flatter gradient.
In fact, radial migration has been proposed as a common phenomenon in the Milky Way history, 
and models exploring this characteristic have been able to explain several chemical 
evolution constraints, such as the metallicity distribution and radial gradients (cf. 
Sch\"onrich \& Binney \citeyear{schonrich}).

\subsection{The Disk Sample divided into Height Groups}

The adopted groups were defined taking steps of 1,000 pc, 600 pc, 500 pc, 400 pc, and 200 pc, 
as follows: 2 groups ($\Delta = 1,000\,$pc, $800\,$pc, and $600\,$pc); 3 groups ($\Delta = 
500\,$pc); 4 groups ($\Delta = 500\,$pc); 6 groups ($\Delta = 400\,$pc), and 11 groups 
($\Delta = 200\,$pc).  A detailed description of the adopted groups is given in Table~2.

\begin{table*}
\small
\caption[]{Definition of the height groups.}
\label{table2}
\begin{flushleft}
\begin{tabular}{ccc}
\noalign{\smallskip}
\hline\noalign{\smallskip}
Number of groups  & Group  & height (pc)      \\
\noalign{\smallskip}
\hline\noalign{\smallskip}
2  & 1  & $\vert z \vert\leq 1,000$           \\
   & 2  & $\vert z \vert> 1,000$              \\
\noalign{\smallskip}
2  & 1  & $\vert z \vert \leq 800$            \\
   & 2  & $\vert z \vert  > 800$              \\
\noalign{\smallskip}
2  & 1  & $\vert z \vert \leq 600$            \\
   & 2  & $\vert z \vert > 600$               \\
\noalign{\smallskip}
\hline\noalign{\smallskip}
3  & 1  & $\vert z \vert \leq 500$            \\
   & 2  & $1000 \geq \vert z \vert > 500$     \\
   & 3  & $\vert z \vert > 1,000$             \\
\noalign{\smallskip}
\hline\noalign{\smallskip}
4  & 1  & $\vert z \vert \leq 500$            \\
   & 2  & $1,000 \geq \vert z \vert > 500$    \\
   & 3  & $1,500 \geq \vert z \vert > 1,000$  \\
   & 4  & $\vert z \vert > 1,500$             \\
\noalign{\smallskip}
\hline\noalign{\smallskip}
6  & 1  & $\vert z \vert \leq 400$            \\
   & 2  & $800 \geq \vert z \vert > 400$      \\
   & 3  & $1,200 \geq \vert z \vert > 800$    \\
   & 4  & $1,600 \geq \vert z \vert > 1,200$  \\
   & 5  & $2,000 \geq \vert z \vert > 1,600$  \\
   & 6  & $\vert z \vert > 2,000$             \\
\noalign{\smallskip}
\hline\noalign{\smallskip}
11 & 1  & $\vert z \vert \leq 200$            \\
   & 2  & $400 \geq \vert z \vert > 200$      \\
   & 3  & $600 \geq \vert z \vert > 400$      \\
   & 4  & $800 \geq \vert z \vert > 600$      \\
   & 5  & $1,000 \geq \vert z \vert > 800$    \\
   & 6  & $1,200 \geq \vert z \vert > 1,000$  \\
   & 7  & $1,400 \geq \vert z \vert > 1,200$  \\
   & 8  & $1,600 \geq \vert z \vert > 1,400$  \\
   & 9  & $1,800 \geq \vert z \vert > 1,600$  \\
   & 10 & $2,000 \geq \vert z \vert > 1,800$  \\
   & 11 & $z > 2000$    \\
\noalign{\smallskip}
\hline
\end{tabular}
\end{flushleft}
\end{table*}

In view of the relatively small size of the sample considered, it is unlikely that a 
division into many groups would produce meaningful results, since the number of objects 
at each height would be small, so that we will focus on the division into 2 groups.

\begin{table*}
\small
\caption[]{Coefficients of the linear fits adopting two height groups.}
\label{table3}
\begin{flushleft}
\begin{tabular}{ccccccc}
\noalign{\smallskip}
\hline\noalign{\smallskip}
    & $A$  & $\sigma_A$ & $B$ & $\sigma_B$  & $r$ & $\sigma_r$ \\
\noalign{\smallskip}
\hline\noalign{\smallskip}
$\Delta = 1,000\,$pc & & & & & & \\
Group 1:  $n = 218$ & 8.7773 & 0.0395 & $-$0.0248 & 0.0048 & $-$0.3300 & 0.2600 \\
Group 2:  $n = 45$  & 8.6086 & 0.0756 & $-$0.0172 & 0.0073 & $-$0.3400 & 0.2100 \\
 & & & & & & \\
$\Delta = 800\,$pc & & & & & & \\
Group 1:  $n = 202$ & 8.7790 & 0.0424 & $-$0.0253 & 0.0053 & $-$0.3200 & 0.2600 \\
Group 2:  $n = 61$  & 8.6755 & 0.0661 & $-$0.0209 & 0.0065 & $-$0.3900 & 0.2400 \\
 & & & & & & \\
$\Delta = 600\,$pc & & & & & & \\
Group 1: $n =168$ & 8.7639 & 0.0480 & $-$0.0218 & 0.0059 & $-$0.2800 & 0.2500 \\
Group 2: $n = 95$ & 8.7302 & 0.0529 & $-$0.0268 & 0.0057 & $-$0.4400 & 0.2600 \\
\noalign{\smallskip}
\hline
\end{tabular}
\end{flushleft}
\end{table*}

   \begin{figure}
   \centering
   \includegraphics[angle=0, width=11.0cm]{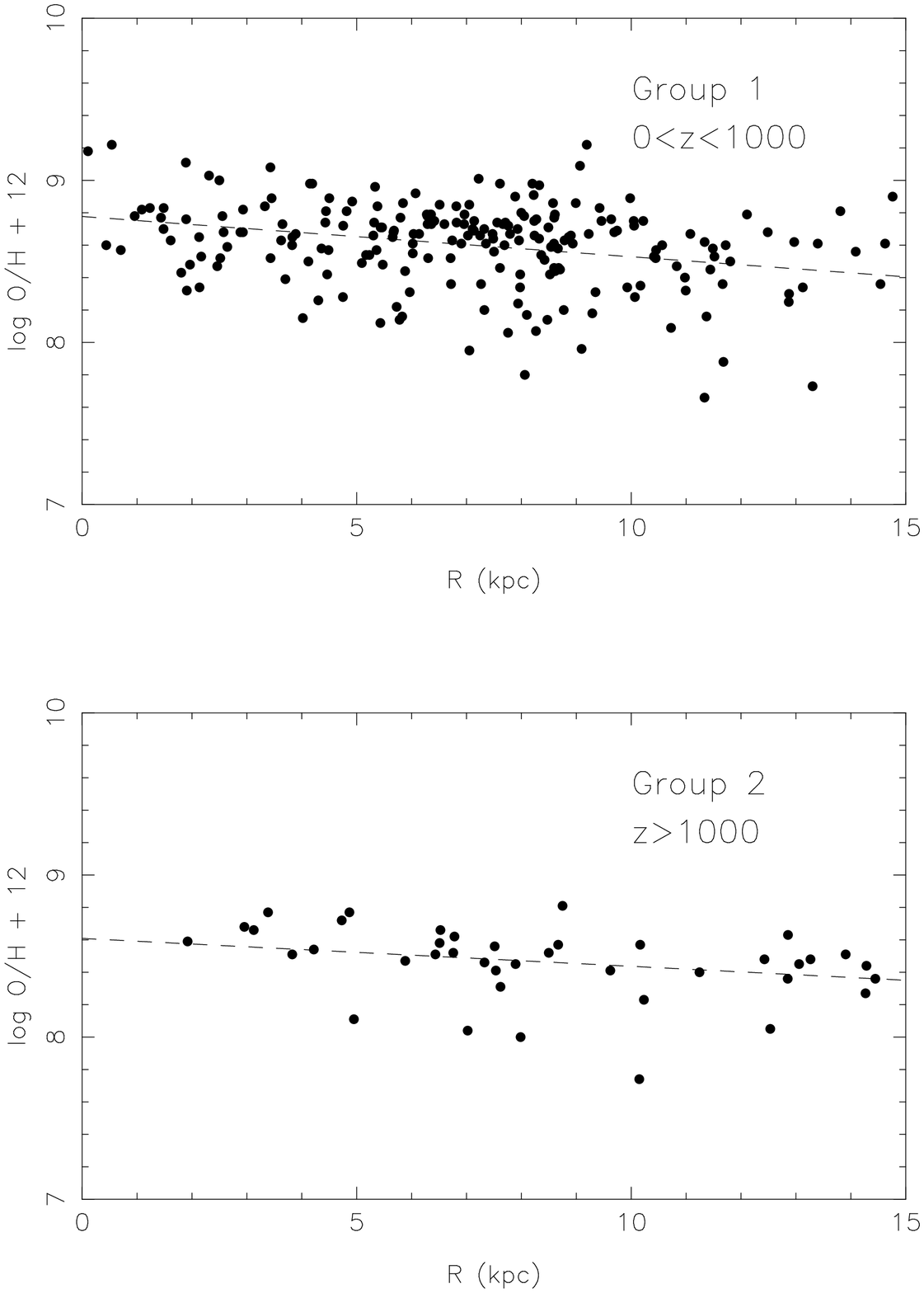}
   \caption{The same as Figure~2, adopting 2 height groups with $\Delta = 1,000\,$pc, Case A.}
   \label{fig4}
   \end{figure}

Let us consider first what is probably the most meaningful result, namely, the division into 
2 groups, with steps of 1,000 pc, that is Group~1 has $\vert z \vert \leq 1,000\,$pc and Group~2 
has $\vert z \vert > 1,000\,$pc. This limit gives an approximate separation of the thin disk and 
the thick disk. The main results are given in the first two rows of Table~3 (Case A), and in 
the plots shown in Figure~4. We can derive the following conclusions:

1. The intercept of Group~1 is higher than that of Group~2, meaning that the abundances of Group~1 
are somewhat higher than those of Group~2, as expected if the former is younger than the latter. 
It can also be seen that Group~2  tends to have lower abundances than Group~1.

2. The gradient of Group~1 is slightly steeper than that of Group~2, but if the uncertainties are taken 
into account both gradients are essentially the same, that is, it is probably not possible to conclude 
that the objects at lower heights have different  gradients from those at higher $z$ values. Again,
the  correlation coefficients are small.

3. There are no evidences of any gradient inversion at low galactocentric distances. It should be 
mentioned  that the samples considered so far do not include objects belonging to the galactic bulge 
(see section~4).

Considering now the remaining divisions into two groups, as given in the remaining rows of Table~3
and in Figures~5 and 6, it can be seen that conclusions~1 and 3 above are still valid, but there is
some difference from conclusion~2 for the groups having $\Delta  = 600\,$pc, in the sense 
that the gradient of Group~2 is slightly steeper than that of Group~1. However, the difference is 
again small, so that conclusion~2 above is also valid for both groups.

   \begin{figure}
   \centering
   \includegraphics[angle=0, width=11.0cm]{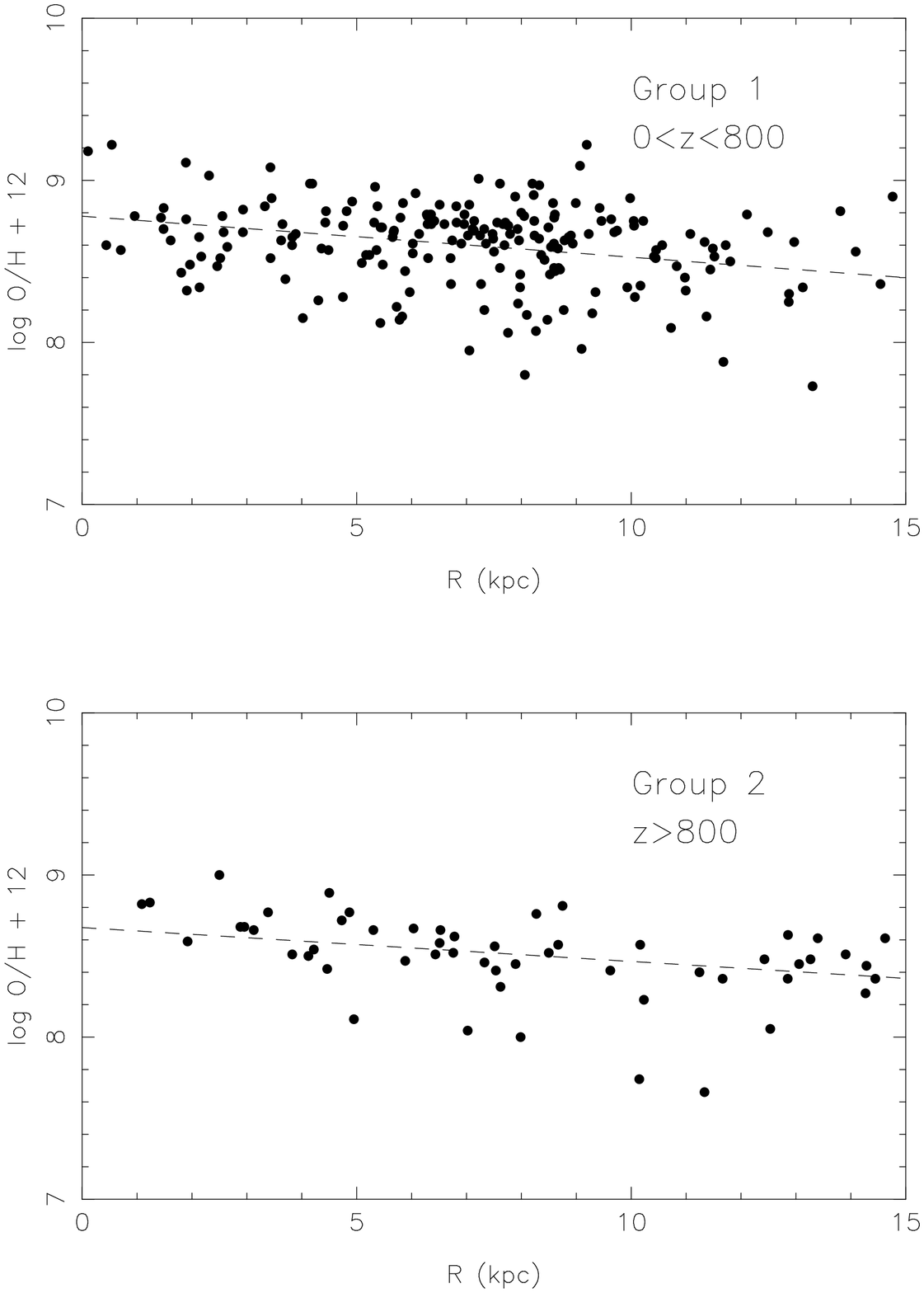}
   \caption{The same as Figure~2, adopting 2 height groups with $\Delta = 800\,$pc, Case A.}
   \label{fig5}
   \end{figure}

   \begin{figure}
   \centering
   \includegraphics[angle=0, width=11.0cm]{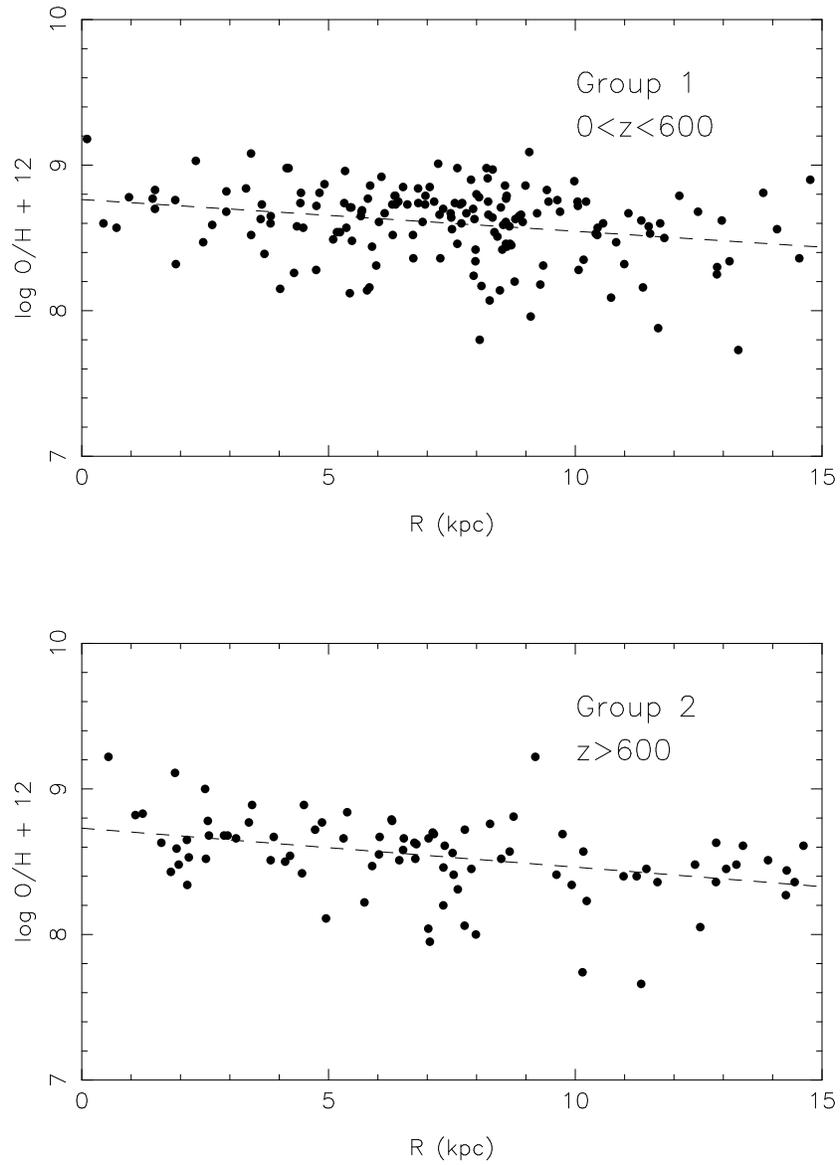}
   \caption{The same as Figure~2, adopting 2 height groups with $\Delta = 600\,$pc, Case A.}
   \label{fig6}
   \end{figure}

The division into 3 groups is the same as in  the previous case for $\Delta = 1,000\,$pc, except 
that Group~1 is further  subdivided into 2 groups. The results are similar, but clearly the new 
Group~2 has relatively few objects compared with the new Group~1. The division into 4 groups is 
also the same, except that both groups are further subdivided into 2 groups. The main difference 
from the previous results is that the new Group~3 has very few objects, thus showing no gradient 
at all. For the division into 6 groups we notice that Groups~4 and 5 have now very few objects, 
so that their gradient is meaningless. Otherwise, the results are similar as in the previous cases. 
For the division into 11 groups, Groups~6 to 10 have very few objects, so that their gradients  
are also meaningless. Otherwise, the results are similar as in the previous cases.

The results of section~4.1 are reinforced by the oxygen and neon gradients determined by
Stanghellini and Haywood (\citeyear{SH}), where different estimates for Peimbert
Type I, II, and III objects have been made. This procedure implicitly assumes that
the gradients are derived at different epochs, since Type I, II and III objects
are expected to reflect increasing ages of the progenitor stars. It is found that
the gradient is slightly flatter for type III nebulae, which are in principle
located higher from the galactic disk, although the differences in the gradients 
are small and similar to those found in the present work. Comparing the results by
SH with the present results for objects at different heights from the Galactic plane,
it can be seen that our gradients at high $\vert z \vert$\ are typically of 
$-0.017\,$dex/kpc or slightly steeper, which is very similar to the Type~III gradients
derived by SH, namely $-0.011\,$dex/kpc. For the objects closer to the disk, we have
typically a gradient of $-0.025\,$dex/kpc, while SH derives $-0.035\,$dex/kpc for Type~I
nebulae, and for the intermediate mass population the corresponding gradients are
about $-0.021\,$dex/kpc, closer to the SH values of $-0.023\,$dex/kpc for Type~II objects.
Therefore, this is a confirmation that the Peimbert types as originally defined by
Peimbert (\citeyear{peimbert1978}) with a few posterior updates reflect the main population
characteristics of the PN in the Galaxy.

\subsection{The Extended Sample: Disk, Bulge, and Interface Region}

The previous samples include essentially objects in the galactic disk, but there are many PN known 
to be in the galactic bulge or in the interface region. The distinction is not always clear, especially 
because of the uncertainties in the distances. Therefore, it is interesting to include these objects 
in our analysis, although some care must be taken to distinguish them from the previous samples of 
disk planetary nebulae.

Our own IAG sample includes 179 objects that in principle belong to the bulge population or to the interface
between the bulge and the disk (see for example Cavichia et al. \citeyear{cavichia1}, \citeyear{cavichia2},
\citeyear{cavichia3}, \citeyear{cavichia4}). 94 objects from this sample  have distances from Stanghellini 
et al. (\citeyear{stanghellini}). The abundances are from the IAG sample, as these objects are not included
in the SH sample. Considering these objects, we have a total sample of 263 + 94 = 357 nebulae.

   \begin{figure}
   \centering
   \includegraphics[angle=0, width=11.0cm]{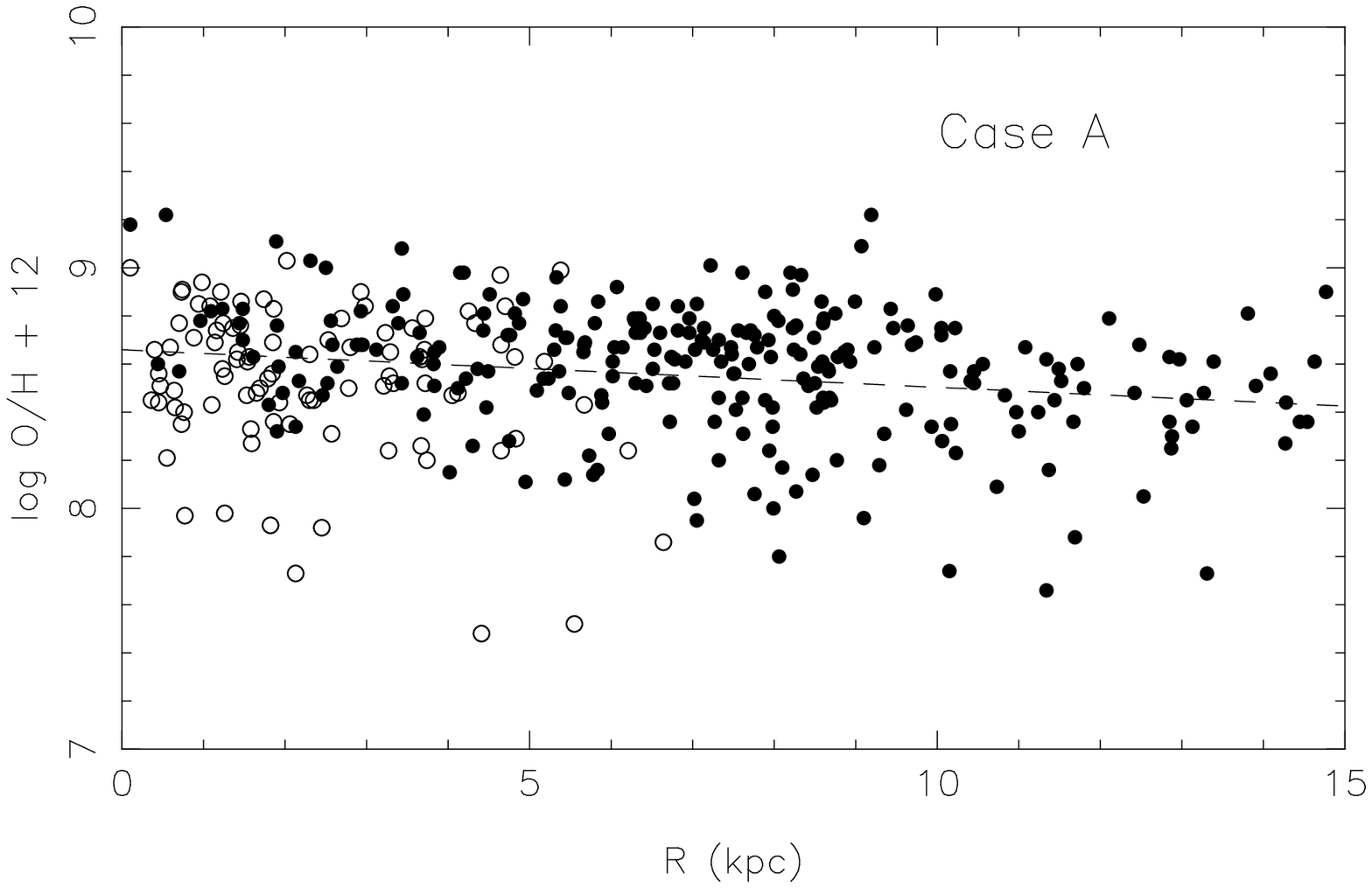}
   \includegraphics[angle=0, width=11.0cm]{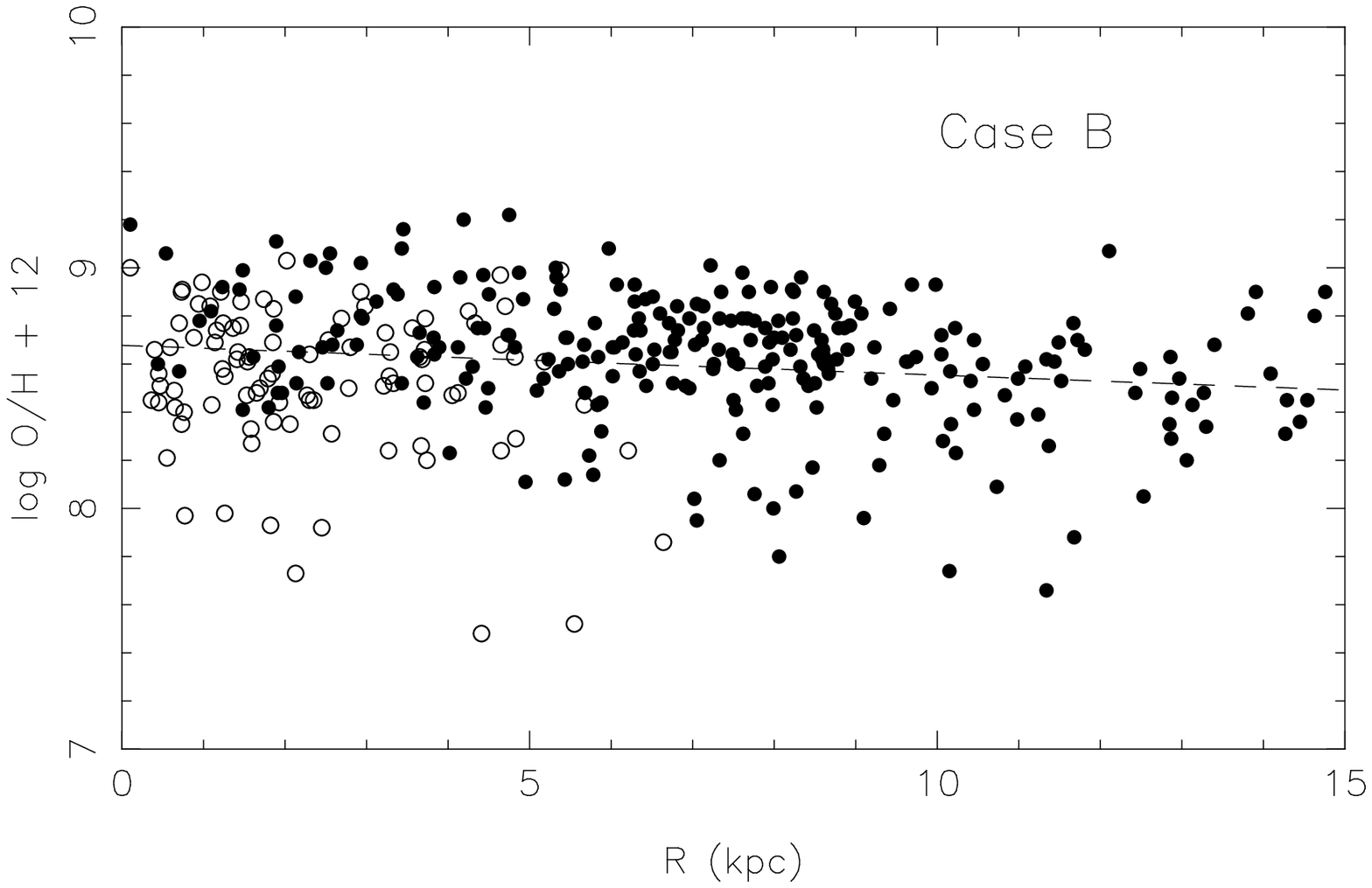}
   \caption{The same as Figure~2, including objects from the bulge and interface region.}
   \label{fig7}
   \end{figure}

The main results are shown in Figure~7, where the filled circles are data for the disk and 
the empty circles for the bulge/interface region. The latter extend to about $R = 6\,$kpc, so 
that some of these objects are clearly not in the galactic centre. It can be seen that the 
inclusion of bulge/interface data slightly decreases the gradients compared with the average 
gradients measured in the outer disk, so that the trend observed for $R > 3-4\,$kpc is somewhat 
modified. This effect should not be confused with a possible gradient inversion at large heights 
$z$ from the plane, since the bulge/interface objects are located very close to the galactic plane.
A similar distribution of bulge nebulae on the O/H $\times \ R$ plane was observed by
Gutenkunst et al. (\citeyear{gutenkunst}). It should be noted that there are some objects  
showing lower abundances ($\log {\rm O/H} + 12 < 8$) than expected by the trend defined by most 
nebulae in both samples. It is possible that these objects can be explained by local abundance 
variations or by the fact that the corresponding central stars are older than most progenitor stars 
in the sample. Also, in this region the galactic bar may play an important role in the shaping of 
the gradient (cf. Cavichia et al. \citeyear{cavichia3}). 

The linear fits are also given in the last two rows of Table~1, again for Cases A and B. The 
slopes are still of the order of $-$0.02 dex/kpc, slightly lower than for the previous cases, 
but they are probably affected by the low abundance objects mentioned above. If we exclude the
outliers, the gradients become closer to the disk sample. We can also obtain a polynomial 
fit to the data as in the previous case. The results are similar, and the gradients vary from a 
minimum of  $-0.014\,$dex/kpc to $-0.037\,$dex/kpc for Case A, with similar results for Case B.

\subsection{Final Remarks}

Conclusions 1-3 listed in section~4.2 are in good agreement with our previous results on the time variation 
of the abundance gradient, as given in Maciel \& Costa (\citeyear{mc2013}).  In particular, conclusion 1 
states the observed differences in the average abundances of PN at high $z$ (older objects) and low $z$ 
(younger objects), and conclusion  2 states the similarity of the gradients of both groups.  The first 
conclusion is consistent with the existence of a vertical abundance gradient, as found in thin and thick 
disk stars of the Milky Way on the basis of stellar data, as can be seen for example in Carrell et al. 
(\citeyear{carrell}) and Chen et al. (\citeyear{chen}). Maciel \& Costa (\citeyear{mc2013})  have 
considered four samples of galactic PN for which the ages of their progenitor stars were estimated 
using three different methods. It was concluded that the younger objects have similar or somewhat 
higher oxygen abundances compared with the older objects, but the gradients are similar within the 
uncertainties. The actual magnitudes of the gradients are in the range $-$0.03 to $-$0.07 dex/kpc with 
an average of $-$0.05 dex/kpc, but depend on the adopted sample. The results of the present paper  
are closer to the lower limit, which is probably a consequence of the adopted distance scale and
the effects of radial migration. Conclusion~2 is also supported by the results by Henry et al.
(\citeyear{henry}), who have separated their planetary nebula sample into two groups adopting
the limit $z = 300\,$pc. Again the slope of the group closer to the galactic plane is slightly
steeper, but the distributions of the two subsamples are essentially the same.

Similar conclusions have been reached in our recent work on the abundance gradients as measured in 
symmetric and asymmetric PN (Maciel \& Costa \citeyear{mc2014}). Since asymmetric PN, especially 
bipolars,  are generally considered as younger than the symmetric objects, some difference in their 
gradients should probably be observed for sufficiently large samples. Considering the elements O,  
Ne, S, and Ar, it was concluded that the average abundances of the bipolar nebulae are somewhat 
higher than for non-bipolars for all elements studied, confirming our conclusion 1 above, but no 
important differences were found between the gradients, which are in the range 
$-0.03\,$ to $-0.05\,$dex/kpc for oxygen.  These results are also supported by some recent work 
by Pilkington et al. (\citeyear{pilkington}) and Gibson et al. (\citeyear{gibson}), who concluded 
that there are no appreciable differences in the gradients in the local universe, near zero redshift, 
although steeper gradients are expected at much higher redshifts, beyond the age bracket considered 
in the present paper.

Cheng et al. (\citeyear{cheng1}, \citeyear{cheng2}) have considered [Fe/H] abundances of a large sample 
of main sequence turnoff stars from the SEGUE survey in the region comprising galactocentric distances 
$6 < R {\rm (kpc)} < 16$ and heights $150 < \vert z \vert {\rm (pc)} < 1500$  relative to the galactic 
plane. They find that close to the disk ($\vert z \vert < 1500\,$pc) the Fe gradient is about 
$-$0.06 dex/kpc, while for higher distances from the plane the gradient flatten out,  with a 
negligible slope for $\vert z \vert > 1,000\,$pc. Their sample is limited to objects with 
$R > 6\,$kpc, so that no information is provided about a possible gradient inversion in the inner disk. 
Since the Fe gradient is probably slightly steeper than the oxygen gradient (see a discussion in Maciel 
et al. \citeyear{mcr2013}), it can be concluded that their results are generally in agreement with our 
present results. 

Acknowledgements. We are indebted to an anonymous referee for some interesting comments and
suggestions on a previous version of this paper. This work was partially supported by FAPESP and CNPq.

\end{document}